\documentclass{article}

\begin{document}
\begin{center}
{\Large\bf Gravitational energy-momentum in small regions according
to the tetrad-teleparallel expressions}
\end{center}

\begin{center}
Lau Loi So\footnote{e-mail address: s0242010@webmail.tku.edu.tw,\\
 present address: Department of Physics, Tamkang University,
Tamsui, Taiwan} and James
M. Nester\footnote{e-mail address: nester@phy.ncu.edu.tw}\ \\
$^{1}$Department of Physics, National Central University,
Chung-Li 320, Taiwan.\\
$^{2}$Department of Physics, Institute of Astronomy, and Center for
Mathematics and Theoretical Physics, National Central University,
Chung-Li 32054, Taiwan.
\end{center}

\begin{abstract} The gravitational energy-momentum within a small region as determined by two tetrad-teleparallel expressions
 is evaluated with the aid of an orthonormal frame adapted to
Riemann normal coordinates.  We find that the gauge current
``tensor'' does enjoy the highly desired and rare property of being
a positive multiple of the Bel-Robinson tensor, whereas M{\o}ller's
expression does not.
\end{abstract}

\bigskip

\section{Introduction: energy-momentum localization}

The localization of energy-momentum for gravitating systems is still
an outstanding fundamental problem \cite{Sza04}.  The classical
attempts to identify a gravitational energy-momentum density for
Einstein's {\em covariant} theory, general relativity, had all led
to various non-covariant expressions which could be written as the
{\em partial derivative} of some particular {\em non-covariant},
coordinate system dependent {\em superpotential}, (see, e.g.,
\cite{Gol58,Tra62,CNC99,Nes04}). As coordinate systems have no
physical significance, these energy-momentum density {\em
pseudotensors} had no clear physical meaning.  This led some to
argue that there was no physically meaningful gravitational
energy-momentum density, and, moreover, that this is just what we
should expect from the equivalence principle (see in particular
\cite{MTW}, \S 20.4).

In 1961 M{\o}ller  constructed an energy-momentum density which,
although itself still a pseudotensor, nevertheless has a
superpotential which is a {\it tensor\/} under coordinate
transformations \cite{Mol61}. M{\o}ller achieved this ``tensor''
form by introducing an orthonormal frame, a tetrad
(a.k.a.~vierbein). His superpotential depends on the local choice of
the orthonormal frame and behaves as a tensor with respect to
coordinate transformations. Like many other energy-momentum
expressions, the value M{o}ller's expression assigns to a spatial
region is not as ambiguous as one might have first thought, it is
{\it quasi-local\/} \cite{Sza04}: it depends on the fields only at
the boundary of the region. More precisely the energy-momentum
M{\o}ller's tetrad expression assigns to a spacetime region
depends---like other pseudotensors---on the boundary choice of the
coordinates, but unlike the other pseudotensors this dependence is
tensorial. Moreover, it also depends on an additional object which
includes non-physical information, namely the choice of tetrad on
the boundary.

M{\o}ller noted that his tetrad description could be given an
interesting reformulation in terms of teleparallel geometry. The
teleparallel reformulation of Einstein's GR (a.k.a. the teleparallel
equivalent of GR (TEGR) and GR$_{||}$) has attracted interest not
only for its presumed advantages for describing energy-momentum but
also as a gauge theory of spacetime translations.  Within the
context of the tetrad-teleparallel theory investigators (see
\cite{Nes81,Nes89,KT91,Sza92,Mal,Maletal99-02,AGP00,Itin02,OR06} and
the works cited therein) have proposed another energy-momentum
expression.  It can be identified as the teleparallel translational
gauge current density.

Nevertheless, largely because of its perceived advantages for
energy-momentum localization, M{\o}ller's tetrad expression
 (even though there is no generally accepted frame gauge
condition)---especially in its interesting teleparallel
description---has continued to attract interest over the years (see,
e.g., \cite{Des63,Mol64,Gol80,KT91,Sza92,BV01} and the works cited
therein).

In certain special cases, however, there is a natural orthonormal
frame; then both M{\o}ller's expression and the gauge current yield
an unambiguous energy-momentum. In particular this is so
asymptotically---at spatial infinity.  In that case M{\o}ller's
expression (like most others) works well (see \cite{Des63} for an
explicit verification; moreover Moller's tetrad expression in fact
also works well at future null infinity \cite{Mol64}). This
asymptotic success is actually not at all surprising; having the
proper asymptotic behavior is a relatively weak requirement, for in
this weak field region an expression need only have the proper
linear theory limit.

The situation is different in the one other situation where there is
a natural frame---a case which has, to our knowledge, not been
previously investigated for the tetrad expressions---namely the
small region limit. In this limit, to zeroth order, one should get
the material energy-momentum density---a quite weak requirement
which follows from the equivalence principle.  On the other hand the
proposed small {\it vacuum\/} region limit is that, to {\it second
order\/}, one gets a positive multiple of the {\it Bel-Robinson
tensor\/} \cite{Gar,DFS99,Sza04} (that would be sufficient to
guarantee that the energy of a small region was {\it positive\/}).
Now this latter requirement is especially interesting as a test of
proposed energy-momentum densities, since it probes the expression
beyond the linear order. It is a strong criterion, capable of
excluding many otherwise acceptable expressions, in particular {\it
none\/} of the classical pseudotensors satisfy this requirement
(although certain artificial combinations of them do
\cite{DFS99,icga7so,So06}).

Here, using Riemann normal coordinates and the associated ``normal''
tetrad, we examine M{\o}ller's expression and the gauge current in
the small region limit. We find that the gauge current {\it
naturally\/} satisfies this highly desirable vacuum Bel-Robinson
property while M{\o}ller's expression does not.

For notation we follow \cite{MTW} unless otherwise noted. Here Greek
indicies are used to refer to spacetime and, unless otherwise noted,
a completely general frame. However, in those sections where it is
necessary to make the distinction, we use Greek indicies to refer to
othonormal frames, with Latin indices reserved for holonomic
(coordinate) frames.

\section{Conserved energy-momentum densities from the field equations}
A gravitational energy-momentum density is easily derived from
Einstein's equations expressed in terms of differential forms:
\begin{equation}
R^\alpha{}_\beta\wedge\eta_\alpha{}^\beta{}_\mu=-2\kappa
T_\mu.\label{eineq}
\end{equation}
Here $\kappa=8\pi G/c^4$ is the gravitational coupling constant (we
will use units with $c=1$), $R^\alpha{}_\beta$ is the curvature
2-form, $T_\mu=T^\nu{}_\mu\eta_\nu$ is the source energy-momentum
3-form, and we are using Trautman's convenient dual form basis
$\eta^{\alpha\dots}:=*(\vartheta^\alpha\wedge\dots)$, where
$\vartheta^\alpha$ is the co-frame. The left hand side of
(\ref{eineq}) is just $-2G^\nu{}_\mu\eta_\nu$, the Einstein tensor
expressed as a 3-form.  Using the definition of the curvature 2-form
in terms of the connection one-form and extracting an exact
differential leads to
\begin{eqnarray}\!\!\!\! R^\alpha{}_\beta\wedge\eta_\alpha{}^\beta{}_\mu&:=&
({\rm d}\Gamma^\alpha{}_\beta+\Gamma^\alpha{}_\gamma\wedge\Gamma^\gamma{}_\beta)\wedge\eta_\alpha{}^\beta{}_\mu\nonumber\\
&\equiv&{\rm
d}(\Gamma^\alpha{}_\beta\wedge\eta_\alpha{}^\beta{}_\mu)
+\Gamma^\alpha{}_\beta\wedge {\rm d}\eta_\alpha{}^\beta{}_\mu
+\Gamma^\alpha{}_\gamma\wedge\Gamma^\gamma{}_\beta\wedge\eta_\alpha{}^\beta{}_\mu\label{decomp}\nonumber\\
&\equiv&{\rm
d}(\Gamma^\alpha{}_\beta\wedge\eta_\alpha{}^\beta{}_\mu)
+\Gamma^\alpha{}_\beta\wedge\Gamma^\lambda{}_\mu\wedge\eta_\alpha{}^\beta{}_\lambda
-\Gamma^\alpha{}_\gamma\wedge\Gamma^\gamma{}_\beta\wedge\eta_\alpha{}^\beta{}_\mu,
\label{rearrangeEineq}
\end{eqnarray}
where we have used $D\eta_\alpha{}^\beta{}_\mu=0$, which follows
since the connection is metric compatible and torsion free.  Using
this expansion one can rewrite the Einstein equation (\ref{eineq})
in a neat form (which is remarkably similar to the form used by
Einstein when he was still searching for a good gravity theory
\cite{JR06}):
\begin{equation}
{\rm d}p_\mu=2\kappa{\cal P}_\mu, \label{pattern}
\end{equation}
where the energy-momentum (superpotential) 2-form is
\begin{equation}
p_\mu:=-\Gamma^\alpha{}_\beta\wedge\eta_\alpha{}^\beta{}_\mu,
\label{superpotential}
\end{equation}
and the  current is the {\it total energy-momentum\/} density
(3-form)
\begin{equation}
{\cal P}_\mu:=t_\mu+T_\mu, \label{total}
\end{equation}
which ``automatically'' satisfies the current conservation relation
${\rm d}{\cal P}_\mu=0$ \cite{Nes04}.  This total energy-momentum
current complex includes the (non-covariant) gravitational
energy-momentum density
\begin{equation}
t_\mu:=(2\kappa)^{-1}\left(\Gamma^\alpha{}_\beta\wedge\Gamma^\lambda{}_\mu\wedge\eta_\alpha{}^\beta{}_\lambda
-\Gamma^\alpha{}_\gamma\wedge\Gamma^\gamma{}_\beta\wedge\eta_\alpha{}^\beta{}_\mu\right).\label{gravemdensity}\label{gaugecurrent3}
\end{equation}
According to this prescription the total energy-momentum within a
region is given by
\begin{equation}
P_\mu(V):=\int_V {\cal P}_\mu =(2\kappa)^{-1}\int_{V}{\rm d}p_\mu
\equiv (2\kappa)^{-1}\oint_{\partial V}p_\mu .\label{total}
\end{equation}

The volume integral form would lead one to expect that the value
depends on the quantities and choice of frame throughout the region,
but the closed 2-surface integral shows that the value is {\it
quasi-local\/}. The value is  still {\it non-covariant\/}: it
depends on the choice of frame---but, as we have already pointed
out, only on the choice at (and, through the connection, near) the
boundary.

The 2-surface integrand is
\begin{equation}
p_\mu:=-\Gamma^\alpha{}_\beta\wedge\eta_\alpha{}^\beta{}_\mu\equiv
-\Gamma^\alpha{}_{\beta\gamma}g^{\beta\sigma}\delta^{\tau\rho\gamma}_{\alpha\sigma\mu}{1\over2}
\eta_{\tau\rho}.
\end{equation}
Expanding the components of this compact 2-form expression gives
\begin{equation}
(\Gamma^\rho{}_{\beta\gamma}g^{\beta\tau}-\Gamma^\tau{}_{\beta\gamma}g^{\beta\rho})\delta^\gamma_\mu+
(\Gamma^\gamma{}_{\beta\gamma}g^{\beta\rho}-\Gamma^\rho{}_{\beta\gamma}g^{\beta\gamma})\delta^\tau_\mu+
(\Gamma^\tau{}_{\beta\gamma}g^{\beta\gamma}-\Gamma^\gamma{}_{\beta\gamma}g^{\beta\tau})\delta^\rho_\mu.
\end{equation}
Specializing to the case where the frame is {\it holonomic} this
expression is exactly the  {\it superpotenial\/} found by Freud
 \cite{Freud};  in that case the associated gravitational energy-momentum
density (\ref{gravemdensity}) is the Einstein {\it pseudotensor\/}
3-form. On the other hand one can choose the frame to be {\em
orthonormal}, then {\it these same formal expressions} become the
those of the tetrad-teleparallel translational gauge current
\cite{Nes89,KT91,Mal,Maletal99-02,AGP00}, which---as we will
elaborate on in the next section---are closely related (see
\cite{Gol80}) to those proposed by M{\o}ller \cite{Mol61} in 1961
(by the way, a differential form construction of these expressions
virtually the same as ours was presented some time ago by Wallner
\cite{Wal80}, and similar arguments appear in \cite{Sza92} and
\cite{Itin02}); the noteworthy thing is that these tetrad expression
are {\it tensors\/}---under coordinate transformations. Although
they are completely independent of the choice of coordinates (i.e.,
they are {\em covariant} under coordinate transformations), they do
depend on the choice of tetrad (in this important sense they are
still non-covariant). More specifically the energy-momentum values
they determine are quasi-local: they depend on the choice of tetrad,
but only on the choice at and near the boundary.

\section{M{\o}ller's expression and the gauge
current}\label{expressions}

The traditional metric approach to gravitational energy-momentum had
led to various pseudotensors (see, e.g.,
\cite{Gol58,Tra62,Sza92,CNC99}), none really satisfactory. Then
M{\o}ller \cite{Mol61} replaced the metric by an orthonormal frame
(a.k.a.\ tetrad, vierbein). The resulting formulation admits an
interesting alternate geometric interpretation in terms of
teleparallel geometry
\cite{Mol61,Nes89,KT91,Mal,Maletal99-02,AGP00,BV01,OR06}. It has
been argued that this framework is more suitable for identifying a
good gravitational energy-momentum density. Indeed, using this
approach M{\o}ller put forward his well-known energy-momentum
expression.

Recall that the Einstein pseudotensor can be obtained as the
(Noether) canonical energy-momentum density from the Hilbert scalar
curvature Lagrangian after a certain (non-covariant) divergence has
been removed (which removes all the second derivatives of the
metric):
\begin{eqnarray}
{\cal L}_{\rm E}(g,\partial g)&:=&\sqrt{-g}R-\partial_l(\sqrt{-g}g^{jm}\Gamma^i{}_{jk}\delta^{lk}_{im}),\\
t_{\rm E}^i{}_j&:=&\frac{\partial{\cal L}_{\rm E}}{\partial
\partial_i g_{kl}}\partial_j
g_{kl}-\delta^i_j {\cal L}_{\rm E};
\end{eqnarray}
this is related to the aforementioned Einstein pseudotensor 3-form
by $t_{\rm E}{}_j=t_{\rm E}^i{}_j\eta_j$.
Similarly, one can obtain M{\o}ller's expression by using the tetrad
$e^\alpha{}_i$ (related to the metric by
$g_{ij}=\bar{g}_{\alpha\beta}e^\alpha{}_i e^\beta{}_j$, where $\bar
g={\rm diag}(-1,+1,+1,+1)$ is the Minkowski metric) as a variable
and removing an appropriate divergence which contains all the second
derivatives of the tetrad:
\begin{eqnarray}
{\cal L}_{\rm M}(e,\partial e)&:=&eR-\partial_l(e e_\alpha{}^i
e_\beta{}^j \Gamma^{\alpha\beta}{}_k \delta^{lk}_{ij}),\label{LMol}\\
t_{\rm M}^i{}_j&:=&\frac{\partial{\cal L}_{\rm M}}{\partial
\partial_i e^\alpha{}_k}\partial_j
e^\alpha{}_k-\delta^i_j {\cal L}_{\rm M}.\label{tMol}
\end{eqnarray}
(Here $e:=\det e^\alpha{}_i$, the dual frame satisfies $e_\alpha{}^i
e^\alpha{}_j=\delta^i_j$ and $e_\alpha{}^i
e^\beta{}_i=\delta^\alpha_\beta$, $\Gamma^{\alpha}{}_{\beta
k}=\Gamma^\alpha{}_\beta(\partial_k)$, and Greek and Latin indicies
are transvected using respectively ${\bar g}_{\alpha\beta}$ and
$g_{ij}$.) The associated M{\o}ller 3-form is $t_{\rm M}{}_j=t_{\rm
M}^i{}_j \eta_i$. From this perspective M{\o}ller's expression is
quite natural, namely it is the (Noether) canonical energy-momentum
density associated with the tetrad variable. It should be noted that
exactly this same density can also be obtained from our
considerations in the previous section---simply by formally
replacing $\mu$ by $j$, while keeping all the other indices
referring to the orthonormal frame.  For more on these two closely
related expressions see \cite{KT91,AGP00}.

In sharp contrast to the metric formulation, within the
tetrad/teleparallel formulation investigators
\cite{Mol61,Nes89,KT91,Mal,Maletal99-02,AGP00,BV01,Itin02,OR06} have
been led to only these two (closely related) quasi-local boundary
term expressions for the energy-momentum within a volume $V$:
\begin{equation}P^{\rm G}_\mu(V):=\oint_{\partial V} p_\mu,\qquad P^{\rm M}_j(V):=\oint_{\partial V}
p_j \equiv \oint_{\partial V} e^\mu{}_j p_\mu,
\end{equation}
 respectively, the {\em translational gauge current} and the
M{\o}ller expression \cite{Mol61}. M{\o}ller had pointed out that
his superpotential (which appears here as a 2-form integrand) is
{\em tensorial} (i.e., it transforms homogeneously under a change of
coordinates); however its differential,
\begin{equation}
t_{{\rm M}j}={\rm d}p_j={\rm d}(e^\mu{}_j p_\mu)={\rm
d}e^\mu{}_j\wedge p_\mu+e^\mu{}_j {\rm d}p_\mu,
\end{equation}
 the M{\o}ller tetrad-teleparallel
energy-momentum 3-form, {\em is not a tensor} with respect to
coordinate transformations (as M{\o}ller himself noted)---because of
the factor ${\rm d}e^\mu{}_i$. In contrast, it should be emphasized
that both the translation gauge current superpotential 2-form
$p_\mu$ and its differential,
 the gauge current 3-form (\ref{gaugecurrent3}), {\em are true tensors}---under
changes of {\em coordinates}.

The tetrad theory, however, does have local Lorentz gauge freedom.
The gauge current expressions do depend on the choice of orthonormal
frame, and thus still contain some observer dependent information
mixed in with the physical information in the energy-momentum
expression. Nevertheless one can regard the gauge current expression
as preferable to any of the pseudotensors or M{o}ller's tetrad
expression, since an orthonormal frame is more physical than an
arbitrary choice of coordinates.

Concerning the ambiguity re the choice of frame, it is important to
note that the {\em quasi-local values} depend only on the choice of
frame on the boundary, and not on the choice within the interior of
the region.

It should also be mentioned that, unfortunately, in some earlier
investigations by our group \cite{TN99,icga7so,So06,SN06} we
misidentified the gauge current as the expression of M{\o}ller.
(From our perspective the gauge current is the natural choice, and
we just assumed that was what M{\o}ller had used---without actually
carefully reading his work.  While we can appreciate that his
expression is---from the Noether approach (\ref{LMol}),
(\ref{tMol})---also a natural choice, the coordinate non-covariance
of his energy-momentum density is certainly a liability.)

\section{Riemann normal coordinates and normal tetrads}
To find the energy-momentum within a small region surrounding a
particular point, we look to the 3-forms ${\cal P}_\mu$, ${\cal
P}_i$ expanding them in a power series.  For this purpose we choose
Riemann normal coordinates $x^i$ centered at the selected point. The
Maclauren-Taylor expansion of the holonomic components of the metric
and connection are well known (see, e.g.~\cite{MTW}, \S 11.6):
\begin{equation}
g_{ij}|_0=\bar g_{ij}, \quad \partial_k g_{ij}|_0=0, \quad
3\partial_{kl}g_{ij}|_0=-R_{ikjl}-R_{iljk},
\end{equation}
\begin{equation}
\Gamma^i{}_{jk}|_0=0,\quad
3\partial_l\Gamma^i{}_{jk}|_0=-R^i{}_{jkl}-R^i{}_{kjl}.
\end{equation}
Here $\bar g_{ij}=\hbox{diag}(-+++)$ is the Minkowski metric.  In
the associated ``normal'' orthonormal frame, the coframe
$\vartheta^\alpha=e^\alpha{}_kdx^k$ and connection one-form
$\Gamma^\alpha{}_{\beta k}dx^k$ components take closely related
analogous values:
\begin{equation}
e^\alpha{}_j|_0=\delta^\alpha{}_j,\quad
\partial_k e^\alpha{}_j|_0=0,\quad
6\partial_{kl} e^\alpha{}_j|_0=-R^\alpha{}_{kjl}-R^\alpha{}_{ljk},
\end{equation}
\begin{equation}
\Gamma^\alpha{}_{\beta j}|_0=0,\quad 2\partial_k
\Gamma^\alpha{}_{\beta l}|_0=R^\alpha{}_{\beta kl}
\label{normalframecon}.
\end{equation}
It is readily verified that these values satisfy, to the appropriate
order, the two relations which transform the metric and connection
coefficients between the holonomic and orthonormal frames:
\begin{equation}
g_{ij}=\bar g_{\alpha\beta} e^\alpha{}_i e^\beta{}_j, \quad
e^\beta{}_j \Gamma{}^\alpha{}_{\beta
i}=\Gamma{}^k{}_{ji}e^\alpha{}_k-\partial_i e^\alpha{}_j.
\end{equation}

\section{Small region values}
Here we present the energy-momentum values for small regions
obtained from the expressions mentioned above.  For non-vacuum
regions all of the expressions reduce in zeroth order to the
material energy-momentum density, in accord with the equivalence
principle. The value obtained for vacuum regions using the holonomic
Einstein pseudotensor has long been known \cite{MTW,Gar,DFS99}. To
second order in RNC it is
\begin{equation}
2\kappa {\cal P}^{\rm E}_j=2\kappa t_{{\rm E}j}\simeq
x^kx^l\frac1{3\cdot6}(4B-S)^i{}_{jkl}\eta_i, \label{einsmall}
\end{equation}
where
\begin{equation}
  S_{\alpha\beta\mu\nu}
:=R_{\alpha\mu\lambda\sigma}R_{\beta\nu}{}{}^{\lambda\sigma}
+R_{\alpha\nu\lambda\sigma}R_{\beta\mu}{}{}^{\lambda\sigma} +\frac14
g_{\alpha\beta}g_{\mu\nu}R_{\lambda\sigma\kappa\rho}R^{\lambda\sigma\kappa\rho},
\end{equation}
and
\begin{equation}
B_{\alpha\beta\mu\nu}
:=R_{\alpha\lambda\mu\sigma}R_{\beta}{}^\lambda{}_{\nu}{}^{\sigma}
+R_{\alpha\lambda\nu\sigma}R_{\beta}{}^{\lambda}{}_{\mu}{}^{\sigma}
-\frac12g_{\alpha\beta} R^{\gamma\sigma\delta}{}_\mu
R_{\gamma\sigma\delta \nu}.
\end{equation}
is the celebrated Bel-Robinson tensor.  (This tensor has many
interesting properties, in particular in vacuum---where the
Riemannian curvature reduces to the Weyl curvature---it is totally
symmetric and traceless.)

\subsection{The tetrad-teleparallel gauge current}
For the gauge current, expanding ${\cal P}_\mu$ using Riemann normal
coordinates and the associated normal tetrad gives, to zeroth order
(unsurprisingly) only the source energy momentum density---just as
it should according to the {\it equivalence principle\/}.  In vacuum
regions ${\cal P}_\mu$ reduces to ${t_\mu}$ (\ref{gravemdensity}),
and the leading non-vanishing value---using
(\ref{normalframecon})---appears at the second order:
\begin{eqnarray}
2\kappa{\cal P}_\mu&=
&\Gamma^{\alpha\beta}\wedge\Gamma^\lambda{}_\mu\wedge\eta_{\alpha\beta\lambda}
-\Gamma^\alpha{}_\gamma\wedge\Gamma^{\gamma\beta}\wedge\eta_{\alpha\beta\mu}\\
&\simeq&\frac{x^lx^m}4\left(R^{\alpha\beta}{}_ {li} R^\lambda{}_{\mu
mj}
-\delta^\lambda_\mu R^\alpha{}_{\gamma li}R^{\gamma\beta}{}_{mj}\right)dx^i\wedge dx^j\wedge\eta_{\alpha\beta\lambda}\nonumber\\
&\simeq&\frac{x^lx^m}4\left(R^{\alpha\beta}{}_ {l\sigma}
R^\lambda{}_{\mu m\delta}
-\delta^\lambda_\mu R^\alpha{}_{\gamma l\sigma}R^{\gamma\beta}{}_{m\delta}\right)\delta^{\nu\sigma\delta}_{\alpha\beta\lambda}\eta_\nu\nonumber\\
 &=&\frac{x^lx^m}4\left(2R_{\mu \lambda m \delta}
R^{\nu\lambda}{}_{l}{}^\sigma
-\frac12\delta^\nu_\mu R^{\gamma\sigma\delta}{}_lR_{\gamma\sigma\delta m}\right)\eta_\nu\\
&=&\frac{x^lx^m}4B^\nu{}_{\mu lm}\eta_\nu, \label{tetradsmallvac}
\end{eqnarray}
proportional to the Bel-Robinson tensor.  In this calculation we
have used the vanishing of the Ricci tensor in vacuum and some well
known curvature tensor symmetry properties.

 To see why it is so desirable to get just the Bel-Robinson tensor one can integrate (\ref{tetradsmallvac}) over a
small coordinate sphere in the surface $x^0=0$, using (with
$a,b,\dots=1,2,3$)
\begin{equation}
\int x^a x^b {\rm d}^3x=\frac13\delta^{ab}\int r^2 {\rm
d}^3x=\frac{4\pi}{3\cdot5}\delta^{ab}r^5,
\end{equation}
and the traceless property of the Bel-Robinson tensor to get for the
gauge current energy-momentum
\begin{equation}
P^{\rm G}_\mu\simeq(2\kappa)^{-1}B^0{}_{\mu
ab}\delta^{ab}\frac{4\pi}{3\cdot4\cdot5} r^5=B^0{}_{\mu
00}\frac{4\pi}{5!\kappa} r^5=\frac1{2G}B^0{}_{\mu 00}\frac1{5!}r^5.
\end{equation}
This result is best appreciated when expressed in terms of the
(traceless, symmetric) electric and magnetic parts of the Weyl
tensor, $E_{ab}:=R_{0a0b}$,
$H_{ab}:=\frac12\epsilon_{acd}R^{cd}{}_{0b}$. We then have a value
similar to that in electrodynamics:
\begin{equation}
P_{\rm G}^{\mu}=(P_{\rm G}^0,P_{\rm G}^c)\simeq\frac{r^{5}}{5!G}
\left(\frac12(E_{ab}E^{ab}+H_{ab}H^{ab}),\epsilon^{acb}E_{ad}H^{d}{}_{b}\right);
\end{equation}
hence $P_{\rm G}^\mu$ satisfies an important energy condition: it is
future pointing and non-spacelike since $P_{\rm G}^0\ge|P_{\rm
G}^c|\ge0$.

\subsection{M{\o}ller's expression}

Turning now to M{\o}ller's expression ${\cal P}_j$:
\begin{eqnarray}
2\kappa(T_j+t^{\rm M}_j)&=&{\rm d}p_j={\rm d}(e^\mu{}_j p_\mu)=e^\mu{}_j{\rm d}p_\mu+{\rm d}e^\mu{}_j\wedge p_\mu\nonumber\\
&\equiv& e^\mu{}_j (2\kappa)(T_\mu+t^{\rm
G}_\mu)+\partial_le^\mu{}_j {\rm d}x^l\wedge
(-\Gamma^{\alpha\beta}{}_m {\rm d}x^m)\wedge\eta_{\alpha\beta\mu}.
\end{eqnarray}
To zeroth order this is again the material result one expects in
accord with the equivalence principle.  For small vacuum regions we
find to lowest non-vanishing order
\begin{eqnarray}
2\kappa t^{\rm M}_j&=&2\kappa t^{\rm G}_j+\partial_le^\mu{}_j
(-\Gamma^{\alpha\beta}{}_m)\delta^{ilm}_{\alpha\beta\mu}\eta_i
\\
&\simeq&\frac{x^lx^m}4
B^i{}_{jlm}\eta_i-\frac16(R^\mu{}_{ljn}+R^\mu{}_{njl})x^n (-\frac12
R^{\alpha\beta}{}_{km}x^k) \delta^{ilm}_{\alpha\beta\mu}\eta_i\\
&=&\frac{x^lx^m}4
B^i{}_{jlm}\eta_i-\frac{x^lx^m}{24}(2B+S)^i{}_{jlm}\eta_i\\
&=& x^lx^m\frac{1}{4\cdot6}(4B-S)^i{}_{jlm}\eta_i.
\end{eqnarray}
Remarkably it turns out to be proportional to the Einstein value
(\ref{einsmall}). (As far as we can see this is just an accidental
coincidence.)

According to this measure the energy within a small sphere of radius
$r$ is
\begin{eqnarray}
P^0&=&\frac1{12\kappa}(B-\frac14 S)^{00}{}_{ab}\int x^ax^b {\rm
d}^3x=\frac{4\pi}{12\kappa} (B-\frac14 S)^{00}{}_{ab}\delta^{ab}
\frac{r^5}{3\cdot5}\\
&=&\frac1{3G}\frac1{5!}r^5(7 E_{ab}E^{ab}-3 H_{ab}H^{ab}),
\end{eqnarray}
which can be negative.

\section{Conclusion}
One reason that the tetrad-teleparallel formulation of GR has been
favored is because it has been believed to have some advantage with
respect to the long-standing problem of how to localize
gravitational energy.  Within this framework two energy-momentum
expressions have been advocated.  Here we have shown that the
tetrad-teleparallel gauge current (which had already been recognized
as one of the best descriptions of the gravitational energy-momentum
for GR) satisfies another important criterion.  Whereas the desired
small region Bel-Robinson property
 is {\em not} satisfied by M{\o{ller's
expression, it is {\it naturally\/} satisfied for the
tetrad-teleparallel gauge current energy-momentum density. An
important consequence is that the gravitational energy according to
the latter measure is {\it positive\/}, at least to this order. (We
expected this positivity result since in fact there is a positivity
 proof for the energy associated with the tetrad gauge current expression
\cite{Nes89}.)

We stress that the vacuum small region Bel-Robinson property is, as
exemplified by the two cases considered here, a strong test capable
of excluding many otherwise acceptable expressions;  indeed {\it
none\/} of the classical pseudotensors (and in this category one can
include M{\o}ller's tetrad-teleparallel expression) satisfies this
requirement (although certain quite artificial combinations of them
do \cite{DFS99,icga7so,So06}).

Compared to these the tetrad-teleparallel gauge current
energy-momentum density stands out.  It is certainly a better
description for gravitational energy-momentum. In addition to being
a tensor (under coordinate transformations) it also enjoys the
highly desired and rare property of having its small region value be
a positive multiple of the Bel-Robinson tensor.

\section*{Acknowledgments}
We would like to thank Prof.~S. Deser for a helpful remark, C. M.
Chen, and the NCTS gravity and cosmology focus group for many
stimulating discussions, as well as the Taiwan NSC for their
financial support under the grant numbers NSC 93-2112-M-008-001,
94-2112-M008-038, 95-2119-M008-027 and 96-2112-M-008-005. JMN was
also supported in part by the National Center of Theoretical
Sciences.


\begin{thebibliography}{33}





\bibitem{Sza04} L.~B.~Szabados,
 Living Rev. Relativity
{\bf 7}, 4 (2004), http://www.livingreviews.org/lrr-2004-4.

\bibitem{Gol58} J. N. Goldberg, Phys.~Rev.~{\bf111},  315
(1958).

\bibitem{Tra62} A. Trautman, in {\sl An Introduction to Current
Research}, edited by L. Witten (Wiley, New York, 1962) pp 169--198.

\bibitem{CNC99}
    C.-C. Chang, J. M. Nester, and C.-M. Chen,
    Phys.~Rev.~Lett.~{\bf83} 1897--901 (1999);
     gr-qc/9809040.

\bibitem{Nes04} {J. M. Nester},
 Class.~Quantum
Grav.~{\bf21}, S261--S280 (2004).

 \bibitem{MTW} C. W. Misner, K. Thorne, and J. A. Wheeler, {\sl
Gravitation\/}, (Freeman, 1973).

\bibitem{Mol61} C. M\o ller,
 Ann.~Phys.~{\bf12}, 118--33 (1961);
 Mat.~Fys.~Dan.~Vid.~Selsk.~{\bf1},
No.10,  1--50 (1961).

\bibitem{Nes81} J. M. Nester, Phys.~Lett.~A, {\bf 83}, 241 (1981)



\bibitem{Nes89} J. M. Nester,
 Int.~J. Mod.~Phys.~A {\bf4},  1755--1772
(1989); Phys.~Lett.~A {\bf139}, 112--114 (1989).

\bibitem{KT91} T. Kawai and N. Toma, Prog.~Theor.~Phys.~{\bf85},
90 (1991); T. Kawai, Phys.~Rev.~D {\bf 62}, 10414 (2000).

\bibitem{Sza92} L.~B.~Szabados, {\it Class. Quantum Grav.\/}, {\bf
9}, 2521--41  (1992).


\bibitem{Mal} J. W. Maluf,
 J. Math.~Phys.~{\bf36}, 4242--4247 (1995); gr-qc/9504010.
 J. Math.~Phys.~{\bf37}, 6293--6301 (1996); gr-qc/9505008.

\bibitem{Maletal99-02}
 J. W. Maluf and J. F. da Rocha-Neto, J. Math.~Phys.~{\bf40},
 1490--1503 (1999); gr-qc/9812020;
J. W. Maluf, J. F. da Rocha-Neto, T. M. L. Toribio, and K. H.
Castello-Branco, Phys.~Rev.~D~{\bf65}, 124001 (2002); gr-qc/0204035.

\bibitem{AGP00} V. C. de Andrade, L. C. T. Guillen, and J. G. Pereira,
 Phys.~Rev.~Lett.~{\bf84},  4533--4536 (2000);
    { gr-qc/0003100}.

\bibitem{Itin02} Y. Itin, Class.~Quantum Grav.~{\bf 19}, 173--89
(2002); Gen.~Rel.~Grav.~{\bf 34}, 1819--37 (2002).

\bibitem{OR06} Y. N. Obukhov and G. F. Rubilar, Phys. Rev. D
{\bf73}, 124017 (2006).



\bibitem{Des63} S. Deser, Phys.~Lett.~{\bf7}, 42 (1963).


\bibitem{Mol64} C. M\o ller,
 Mat.~Fys.~Medd.~Dan.~Vid.~Selsk.~{\bf34},
no.3,  1--67 (1964).



\bibitem{Gol80} J. N. Goldberg, ``Invariant Transformations,
Conservation Laws, and Energy-Momentum'', in {\sl General Relativity
and Gravitation: One Hundred Years After the Birth of Albert
Einstein}, Vol 1, ed A. Held (Plenum, New York, 1980) pp 469--489.

\bibitem{BV01} M. Blagojevi\'c and
M. Vasili\'c, Phys.~Rev.~D, 64, 044010 (2001).


\bibitem{Gar} J. Garecki, Acta Phys.~Polon.~B {\bf4}, 537
(1973); Class.~Quantum Grav.~{\bf2}, 403--408 (1985);
Ann.~Phys.~{\bf10}, 911--919 (2001); gr-qc/0003006.



\bibitem{DFS99}
S. Deser, J. S. Franklin, and D. Seminara, Class.~Quantum
Grav.~{\bf16}, 2815--2821 (1999).

\bibitem{icga7so}
 L. L. So,  J. M. Nester, and H. Chen, 2006\hfil\break ``Classical
pseudotenors and positivity in small regions''\hfil\break in {\sl
GRAVITATION AND ASTROPHYSICS} {\it on the occasion of the 90th year
of General Relativity}
 Proceedings of the VII International Conference on Gravitation
and Astrophysics, ed J. M. Nester, C.-M. Chen and J. P. Hsu (World
Scientific, 2006) p 356.


\bibitem{So06} L. L. So,
``Quasi-local energy-momentum and pseudotensors for GR in small
regions'', PhD thesis, (National Central University, 2006),
unpublished.


\bibitem{JR06} M. Janssen and J. Renn, ``Untying the Knot: How
Einstein Found His Way Back to Field Equations Discarded in the
Zurich Notebook'', in {\sl The Genesis of General Relativity} Vol.
2, {\sl Einstein's Zurich Notebook: Commentary and Essays} ed J.
Renn (Springer, 2006) pp 849--925.


\bibitem{Freud} Ph. Freud, Ann.~Math.~{\bf40}, 417--419 (1939).




\bibitem{Wal80} R. P. Wallner, Acta Phys.~Aust.~{\bf52}, 121--4
(1980).



\bibitem{TN99} R.-S. Tung and J. M. Nester, Phys.~Rev.~D {\bf
60}, 021501 (1999).






\bibitem{SN06} L. L. So and J. M. Nester, ``Gravitational
energy-momentum in small regions according to M{\o}ller's tetrad
expression'', gr-qc/0612061 v1.




\end{thebibliography}
\end{document}